# The profile likelihood ratio and the look elsewhere effect in high energy physics

Gioacchino Ranucci

*Istituto Nazionale di Fisica Nucleare*

*Via Celoria 16 - 20133 Milano*

*Italy*

*Phone: +39-02-50317362*

*Fax: +39-02-50317617*

*e-mail: gioacchino.ranucci@mi.infn.it*

The experimental issue of the search for new particles of unknown mass poses the challenge of exploring a wide interval to look for the usual signatures represented by excess of events above the background. A side effect of such a broad range quest is that the significance calculations valid for signals of known location are no more applicable when such an information is missing. This circumstance is commonly termed in high energy physics applications as the look elsewhere effect. How it concretely manifests in a specific problem of signal search depends upon the particular strategy adopted to unravel the sought-after signal from the underlying background. In this respect an increasingly popular method is the profile likelihood ratio, especially because of its asymptotic behavior dictated by one of the most famous statistic result, the Wilks' theorem. This work is centered on the description of the look elsewhere effect in the framework of the profile likelihood methodology, in particular proposing a conjecture about the distribution of the likelihood ratio under the null hypothesis of absence of the signal over the searched mass interval, a condition which is known to violate the conditions of the Wilks' theorem. Several Monte Carlo tests to support the conjecture are presented, as well.

# 1. Introduction

Maximum likelihood ratio tests are surely among the most used methods in statistics [1], generally applicable to a variety of parametric hypothesis testing problems. One of the reasons of their popularity is traced to the well known Wilks' theorem [2], which states that their asymptotic null distributions belong to the $\chi^2$ family functions and are independent of nuisance parameters.

An example of this kind of test that has recently drawn substantial attention in HEP is the profile likelihood ratio [3].

Considering the generic situation of an HEP experiment which produces as output of the measurements properly binned data, the usual convention is to denote the expectation value of the measured data in a generic bin $i$ as

$$E(n_i) = \mu s_i + b_i \qquad (1)$$

where $b_i$ is the background contribution, $s_i$ the signal contribution according to the nominal value of the model, and $\mu$ is the signal strength parameter; therefore $\mu=0$ corresponds to the background-only hypothesis, while $\mu=1$ reproduces the nominal signal hypothesis.

Given a data outcome, one can construct the likelihood function $L$ as the product of Poisson probabilities for all bins

$$L(\mu, \boldsymbol{\theta}) = \prod_{i=1}^{M} \frac{(\mu s_i + b_i)^{n_i}}{n_i!} e^{-(\mu s_i + b_i)} , \qquad (2)$$

which will depend upon the strength parameter $\mu$ and a set of nuisance parameters collectively denoted with $\boldsymbol{\theta}$, typically connected with the shape of both the signal and the background, as well as with the total background rate. The function $L$ could contain also terms related to auxiliary measurements, for example in side regions to constrain the background, not show in (2) for simplicity.

In this framework, purpose of the profile likelihood ratio is to test a hypothesized value of $\mu$ against the alternatives, and it is written as

$$\lambda(\mu) = \frac{L(\mu, \hat{\hat{\boldsymbol{\theta}}})}{L(\hat{\mu}, \hat{\boldsymbol{\theta}})} . \qquad (3)$$

At the numerator of the ratio there is the profile likelihood function [4], in which $\hat{\hat{\boldsymbol{\theta}}}$ is the value of $\boldsymbol{\theta}$ maximizing $L$ for the assumed $\mu$; in other words, $\hat{\hat{\boldsymbol{\theta}}}$ is the conditional maximum-likelihood estimator of $\boldsymbol{\theta}$ and consequently is a function of $\mu$ itself. The denominator, instead, is maximized in an unconstrained way, thus $\hat{\mu}$ and $\hat{\boldsymbol{\theta}}$ are the true maximum likelihood estimators. Hence, by definition, the profile likelihood ratio is comprised between 1, when the hypothesized $\mu$ coincides with $\hat{\mu}$, showing thus great compatibility between the data and the hypothesis, and 0 when instead the assumed $\mu$ is at odd with $\hat{\mu}$, denoting in this way a high degree of incompatibility between the data and the hypothesis.

According to the Wilks' theorem terminology, we define the null hypothesis as the condition in which the tested value of $\mu$ coincides with its true value: then under the null hypothesis, and if some required regularity conditions are satisfied, the theorem ensures that the quantity $t = -2\ln\lambda(\mu)$, usually termed as the log-likelihood ratio and adopted as the actual test statistics for the problem under consideration, is asymptotically distributed according to a $\chi_s^2$ function, whose $s$ degrees of freedom are equal to the difference between the number of maximization parameters at denominator and numerator, thus 1 in the present case.

However, we have not considered up to now explicitly the role played in this scenario by the energy location of the sought-after signal. If the test of the hypothesized $\mu$ occurs for a fixed mass value then the previous considerations remain unchanged, but if instead the energy position of the signal, denoted with $E$, is another unknown of the search process, which corresponds to say that $E$ falls among the nuisance parameters, then things change radically while looking for a discovery.

To be clear on this point it is useful to remind that, as elucidated in [3], the profile likelihood ratio can be employed for a twofold purpose, either to claim a discovery of a new signal or to put upper limits in case of absence of evidence of new physics, exploiting in both cases the features of the distribution of the quantity $t$ under the null hypothesis.

For example, in the case of the upper limit, for which the typical considered confidence level in HEP applications is 95%, a threshold is imposed on the null hypothesis distribution such that its integral from 0 to the assumed threshold is 0.95. Then, given a specific experimental outcome, the quantity $t = -2\ln\lambda(\mu)$ is computed for the actually observed data and for the value of $\mu$ for which the limit is sough-after, being its specific outcome denoted as $t_{obs}$. If $t_{obs}$ is found equal or above the threshold, the corresponding $\mu$ is excluded at a level equal or greater 95%. Alternatively, given the observed data, the same quantity $t = -2\ln\lambda(\mu)$ is computed for varying $\mu$ until it equals the 95% threshold imposed on the null hypothesis distribution: the $\mu$ for which such an equality is reached is the sought-after 95% upper limit.

We focus our attention in this work on the discovery scenario, which means that the specifically tested value of $\mu$ is 0: the goal of the profile likelihood ratio in this context is to define the degree of (in)compatibility of the data with the background only hypothesis, which would lead to a discovery if incompatibility is assessed, at least, at 5 sigma level. Similarly to the upper limit case, operationally this is done defining a threshold on the corresponding null hypothesis distribution in agreement with the desired confidence level. For the typical discovery 5 sigma value advocated in HEP scenarios, this amounts to ensure that the integral of the null hypothesis distribution from the threshold to infinite is 2.87E-7. Upon computing the quantity $t = -2\ln\lambda(\mu)$ for $\mu$=0 and for the actually observed data, if it is found equal or greater than the threshold, then a discovery at 5 sigma level is claimed.

As well known, the quest for new physics performed through this procedure implies the search over some smooth invariance mass distribution of an excess of events, a bump, which could occur anywhere in the investigated mass range, thus we immediately find us in the above mentioned situation in which $E$ is one of the nuisance parameters $\theta$ of the test.

At this point it is important to remind that one of the regularity conditions for the validity of the Wilks' theorem is that the "restricted" parameters, i.e. those that at numerator are kept fixed at the values that one desires to test, must assume values within their admissible range of variability, but strictly not at the boundary. To be more explicit, given the interval of variability for each parameter upon which the denominator maximization is carried out, each tested value at numerator must stay within the corresponding interval but not at the border.

In order to understand better this specific aspect and its subsequent implication in the signal search procedure, let's come back for a while to the situation in which the tested $\mu$ is different from 0, and consider the case in which we want to test a model that predicts not only the value of the strength parameter, but also that of *E*. It is thus convenient to make E explicit in the formulation of the profile likelihood ratio, which becomes

$$\lambda(\mu, E) = \frac{L(\mu, E, \hat{\hat{\boldsymbol{\theta}}})}{L(\hat{\mu}, \hat{E}, \hat{\boldsymbol{\theta}})} \tag{4}$$

If, as it would be extremely likely, the tested $\mu$ and *E* are well within the respective admissible ranges of variability, i.e. the ranges considered in the denominator maximization, then under the null hypothesis that the tested model were true, the quantity $t = -2\ln\lambda(\mu, E)$ is distributed according to a $\chi_s^2$ distribution with *s*=2 degrees of freedom.

This nice picture is spoiled, on the other hand, when we deal with the situation we are actually interested to, i.e. the test of the background only condition against the possible presence of a signal anywhere in the energy range.

Since things become troublesome with respect to both parameters $\mu$ and *E*, let's first consider the case of $\mu$. Incompatibility of the background only hypothesis with the data may happen either for largely positive or largely negative values of the strength parameter. Thus, from a pure mathematical standpoint, $\mu$ can be both positive and negative; therefore if we fix at numerator $\mu$=0 the requirement of the Wilks' theorem to test a value within its admissible range is fulfilled. Obviously, whether to allow $\mu$ to be negative is physically meaningful is a separate issue; one may, however, note that there are experimental occurrences in which even the detection of less than expected events is a sign of an effect, like in neutrino oscillation experiments. On the other hand, this is not the case for the search of new physics at colliders, where instead by definition the signal constitutes a positive effect. Anyhow, initially the occurrence that the strength parameter takes on both positive or negative values is considered, thus fulfilling the regularity conditions with respect to $\mu$, nevertheless with the additional constrain that in any case the sum of signal and background is bounded to be non negative. In particular all the calculations in paragraph 4 are carried out in this condition, while the discussion of the positive-only case is postponed to paragraph 5.

Rather, the situation is more drastic for the parameter *E*, which simply does not exist in the background only condition that we want to test; actually, it exists only under the alternatives (i.e. at denominator) and this causes a breakdown of the validity conditions of the Wilks' theorem that cannot be remedied .

Therefore when using the discovery test statistics $t_0$ defined as

$$\lambda(\mu = 0, \hat{E}) = \frac{L(\mu = 0, \hat{\hat{\boldsymbol{\theta}}})}{L(\hat{\mu}, \hat{E}, \hat{\boldsymbol{\theta}})} \tag{5}$$

$$t_0 = -2\ln\lambda(\mu = 0, \hat{E}) \tag{6}$$

in the search of a new signal which can appear anywhere in a mass range, its distribution cannot be obtained resorting to the Wilks' theorem.

It is interesting to note that the test statistics $t_0$ for the discovery of a new signal as written in (5) and (6) effectively amounts to test individually each energy value, assuming then as test statistics the maximum among all of the computed log-likelihood ratios (which occurs at the energy denote with $\hat{E}$). To stress this specific aspect of the signal discovery problem, such a scenario is expressively termed in HEP as Look Elsewhere Effect (LEE) [5] , a terminology that outlines the fact that the signal is searched over a broad range, and not in an a-priori defined energy location. Quantitatively, the main practical implication of the LEE is that the significance of a signal detection is substantially altered with respect to a standard search at a fixed mass value, so that usually when referring to this effect one implicitly means its impact in term of significance of a signal detection.

## 2. Conjectured form of the distribution of the log-likelihood ratio test statistics

The specific operational importance of the distribution of the test statistics $t_0$ relies on the fact that the integration of its tail gives the *p-value* (and hence the significance) for the observation of a putative new signal in the explored mass range. Specializing what expressed in general in the previous paragraph, incompatibility with the background only hypothesis leading to a potential discovery implies values of $\lambda(\mu=0,\hat{E})$ close to 0, while compatibility with the same hypothesis is signaled by values close to 1. In terms of the test statistics $t_0$ this means that growing values of it are obtained when the data are more and more incompatible with the background only hypothesis, while values close to 0 correspond to the acceptance of the background hypothesis.

Quantitatively one can set a discovery threshold on the tail of the distribution of $t_0$ so to ensure the desired significance of a possible signal detection. For example, a 5 sigma significance detection means to set a threshold such that the probability to find greater values of $t_0$ in the occurrence of absence of a signal is less than 2.87 x $10^{-7}$. Therefore, since the interest in a discovery process is for very low *p-value*, it can be enough for this purpose to know only the extreme tail of the distribution of the test statistics $t_0$.

Obviously the tail of $t_0$ is different, essentially enhanced, with respect to what would have been in case of detection at fixed mass, therefore we can regard such a modification of the $t_0$ tail as the concrete way in which the Look Elsewhere Effects comes into play while searching for a new signal through the profile likelihood ratio approach. Anyhow, this consideration about the "important" part of the distribution, e.g. the tail, should not hide that actually the shape of the entire $t_0$ distribution is altered by the LEE, in particular deviating in the example discussed here from the Wilks predicted $\chi_2^2$ behavior.

Very relevant to the present discussion is the result reported in reference [6], where an upper bound is given for the *p-content* of the tail of $t_0$ under the assumption that its corresponding distribution for fixed $E$ is $\chi_s^2$ distributed with *s* degree of freedom. In reference [7] the authors show how this general result can be practically employed for an explicit expression of the tail of the $t_0$ distribution stemming from the LEE effect, with the help of a procedure based on a low statistic Monte Carlo. Interestingly, the authors of [7] also show that the mathematics exploited to get the asymptotic expression of the tail suggests an heuristic interpretation of the procedure, in which the investigated mass range can be considered as subdivided in *N* effectively independent search regions, in each of which the log-likelihood ratio $t_0$ locally fluctuates as a $\chi^2_{s+1}$ distribution with *s+1* degree of freedoms. Essentially, it is as if the *s* degrees of freedom for $t_0$, while the search is carried out at fixed *E*, were increased in each "local" search region of 1 to account for the extra variability associated to the indetermination of the mass position.

Therefore, in the framework of this effective interpretation, the search for the maximum of the log-likelihood ratio $t_0$ over the energy interval of interest is equivalent to identify the largest among *N* independent $\chi^2_{s+1}$ variables. For completeness, the search procedure includes also an extra $\chi^2_s$ variable, which is introduced to account for the possibility of the occurrence of the global maximum at the boundary of the search region: in this case the mass location is fixed and hence the extra 1 degree of freedom should not be considered .

The authors of [7] finally remarks how their interpretation of the search algorithm as applied to *N* independent sub-mass ranges can be naturally reconciled with the description of the LEE reported in [8].

Obviously the key factor to transform this interpretation into a concrete evaluation is the determination of *N*, and actually the low statistic MC procedure illustrated in [7] is devoted just to this purpose.

Building on these previous achievements, in this work it is further conjectured that the entire, true distribution (PDF) of the test statistic $t_0$ can be approximated through a proper combination of $\chi^2_{s+1}(t_0)$ and $\chi^2_s(t_0)$ functions as follows:

$$p(t_0) = N \int_0^{t_0} \chi^2_s(h)dh \left( \int_0^{t_0} \chi^2_{s+1}(h)dh \right)^{N-1} \chi^2_{s+1}(t_0) + \chi^2_s(t_0) \left( \int_0^{t_0} \chi^2_{s+1}(h)dh \right)^N \qquad (7)$$

In this formula, according to the above terminology, *N* is the effective number of independently searched energy regions, in which $t_0$ is locally $\chi^2_{s+1}(t_0)$ distributed, while the $\chi^2_s(t_0)$ terms arise from the possibility of the maxima at the two extremes of the allowed energy range.

It is worth to point out that this model is inspired by the concept of "number of effectively scanned frequencies", with the associate mathematical framework, which arises in the context of the search of modulations of unknown periods possibly embedded in time series data [9] [10] . A thorough account of the parallelism of this approach with the search of a particle of unknown mass procedure has been given in [11].

When $t_0$ is large, the integrals in (7) are all approximately equal to 1, and thus asymptotically (7) reduces to

$$p(t_0) = N\chi^2_{s+1}(t_0) + \chi^2_s(t_0) \qquad (8)$$

e.g. the asymptotic tail approximation reported in [6][7], which is thus correctly recovered in the conjectured complete model for the $t_0$ PDF.

Expression (7) stems from the application of order statistics [12] to the case under consideration: we have indeed *N+1* terms fluctuating independently and we thus simply write the probability density function of the highest among them using the prescription of order statistics when the terms are independent each other. Specifically, the first addend in (7) describes the occurrence that the largest term is one of the $N$ $\chi^2_{s+1}(t_0)$ distributed variables, thus the fourth factor in the first addend describes the probability density of the largest variable, the two integrals on its left express the probability that no one of the remaining variables (*N-1* $\chi^2_{s+1}(t_0)$ distributed and one $\chi^2_s(t_0)$ distributed) exceed the value of the largest, and *N* is inserted to account for the number of possible choices of the largest among the *N* variables $\chi^2_{s+1}(t_0)$ distributed .

The second addend in the sum, similarly, represents the probability that the largest variable is the one $\chi^2_s(t_0)$, and therefore the integral on its right side expresses the probability that all the N variables $\chi^2_{s+1}(t_0)$ distributed do not exceed its value.

**3. Toy models for numerical calculations**

A mathematical demonstration of the validity of expression (7) appears to be a daunting task. Rather, the approach used in this work is to support the conjecture, as well as the "building blocks" upon which is based, with a set of MC tests carried out on two toy models. The first model is the same adopted in [7]: in a mass range from 0 to 120 a background distribution of Rayleigh type is assumed. The search for a signal is accomplished by hypothesizing for it a Gaussian function of unknown amplitude and location, and width linearly increasing with the mass value, so to reproduce a standard situation of variable resolution as function of the energy.

The second model, instead, is taken from the recently proposed problem of the Banff challenge 2a [13], and it is based on a exponential background in a mass range from 0 to 1, explored for a Gaussian type mass bump of fixed width (σ=0.03).

Therefore in both models the only other nuisance parameter, in addition to the energy location of the searched signal, is the amplitude of the background, that is estimated from the data themselves without resorting to auxiliary measurements to constrain it. This also means that in both examples the variable *s* is equal to 1, i.e. the $\chi^2_s$ distribution under the null hypothesis for the search at a fixed mass has only *s*=1 degree of freedom.

Situations where the number of degrees of freedom is more than 1 can be considered as well, for example assuming several independent channels, each characterized by its individual signal strength parameter. But, for simplicity, the MC evaluation reported in the following are all for the case *s*=1.

The numerical experiments are performed generating in each simulation cycle the background counts in each bin according to its assumed characteristics and then performing the maximizations at numerator and denominator of the profile likelihood ratio (3). At the end of each cycle the value of the test statistics $t_0$, i.e. the maximum over the spanned energy range of the log-likelihood ratio, is recorded and put in a histogram.

The extent of validity of the conjecture of § 2 will be checked by comparing the MC distribution of the test statistics $t_0$ obtained in this way with the model. Furthermore, also the other key elements of the model will be simulated and checked against the assumptions on which the conjecture is based, in particular the fixed mass distribution, and the "local" $t_0$ distribution in the effective independent search regions.

The results reported in the next paragraph 4 are computed allowing the signal strength parameter to be positive or negative, with the caveats mentioned above. The subsequent paragraph 5 is instead devoted to the case of restricting $\mu$ to be only positive.

## 4. Numerical results of the toy models when the strength parameter is allowed to be both positive and negative

*4.1 Toy model based on the Rayleigh invariant mass distribution for the background*

The first discussed example is the Rayleigh model for the background. The resolution is considered linearly variable over the entire range, and this is thus assumed as the width of the presumed signal; given the adopted binned likelihood approach, two different conditions will be evaluated, with bin width respectively 1or 3, in order to get hints of possible binning related effects. Furthermore, the intensity of the background is kept fixed throughout the generation cycles.

The first experimental outcome of the $t_0$ distribution, obtained with 10000 simulations, is reported in Fig. 1 for bin width 1.

By fitting the model (7) to the obtained MC distribution, one can observe that it reproduces remarkably well the simulation data for N=4.48+/-0.05. The agreement is very good also for the tail, as it is better appreciated through the log plot reported in Fig. 2.

With 10000 simulations the only parameter of the model $N$ is constrained very precisely, but following [7] it can be checked if an acceptable evaluation of $N$ can be obtained with a limited number of simulations.

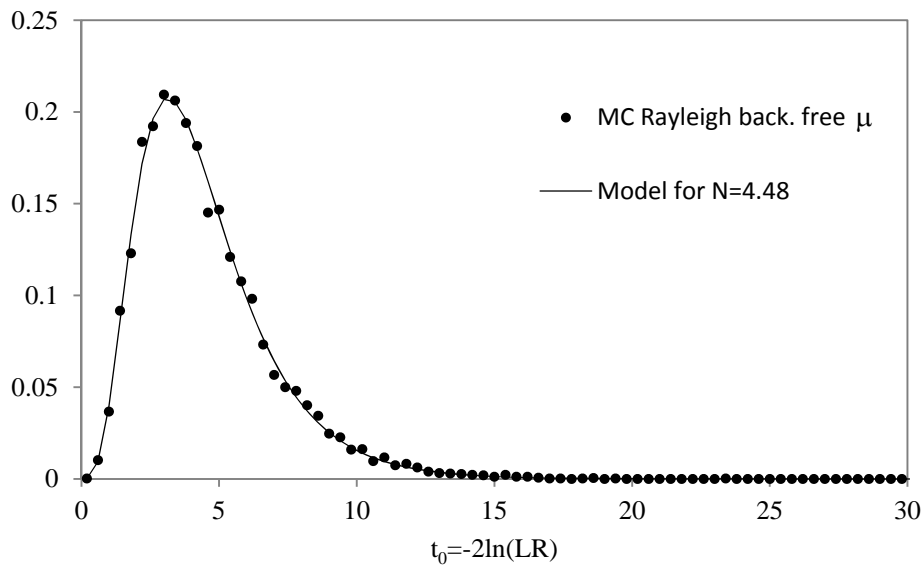

*Figure 1 – Distribution of the $t_0$ test statistics for Rayleigh background and with the signal strength parameter free to assume positive or negative values.*

It comes out that a powerful aspect of the technique is that N can be rather precisely evaluated also with a very low statistic MC, as proved by a test performed with only 100 cycles, from which N has been evaluated to be 4.49 +/-0.5. In passing, one can note how the statistical uncertainty in the determination of N scales exactly as the number of simulation cycles.

To better study specifically the behavior of the tail, which in case of significance assessment at several sigma is the most important and sensitive part of the distribution, the simulation has been repeated with 100000 cycles. The result reported in Fig. 3 shows indeed that the model ensures a really faithful reproduction of the tail.

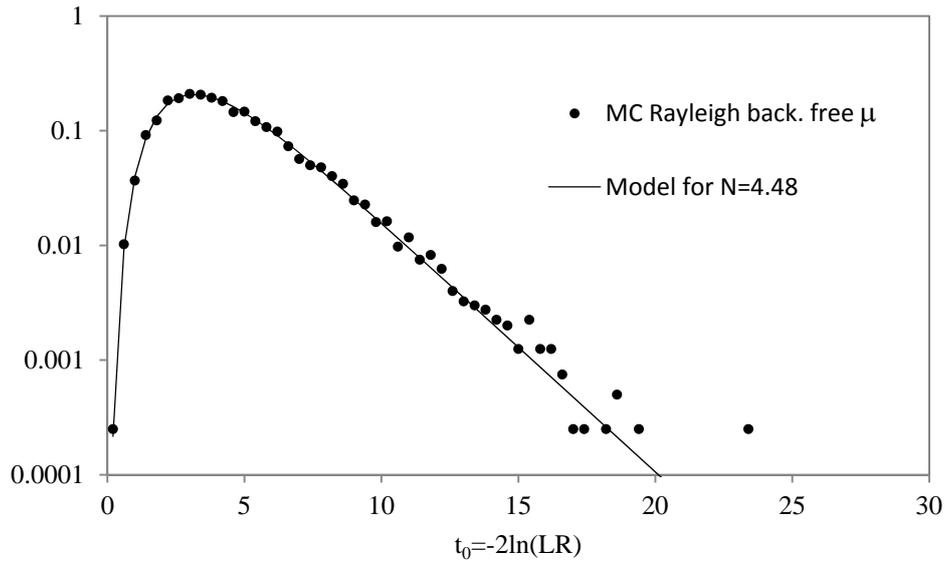

*Fig. 2 – The log plot of the same distribution of Fig. 1 demonstrates the good match of the MC tail with the conjectured*

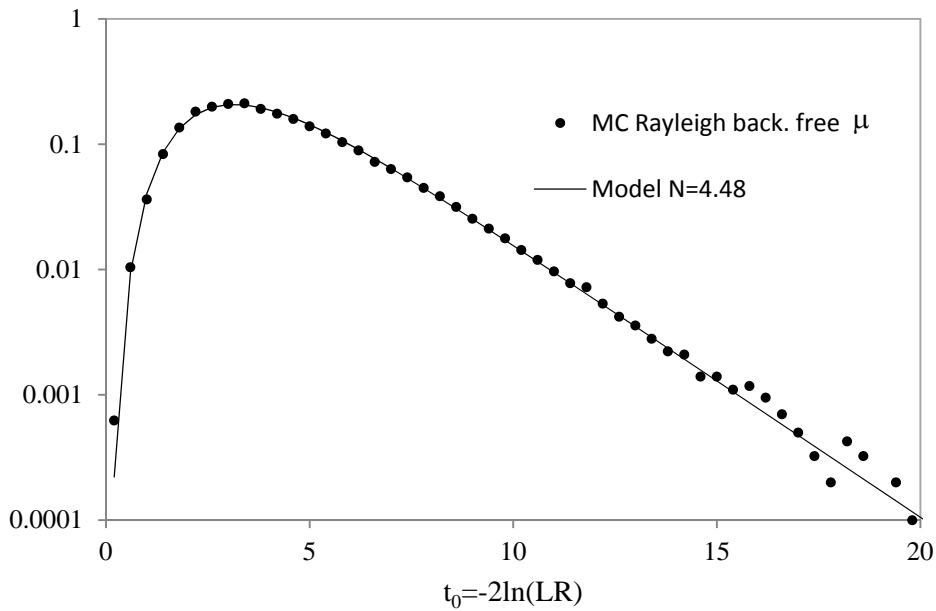

*Figure 3 – High Statistics Monte Carlo showing the exceptional agreement over several orders of magnitude of the tail of the log-likelihood ratio test statistics with the corresponding model*

The result of the test to unravel possible binning-related effects, performed by using three times wide bins, is shown in Fig. 4. The outcome of the simulation is substantially unchanged, showing that the log-likelihood ratio $t_0$ is still well described by the model (7), with the parameter $N$= 4.32 +/- 0.05. Thus, it appears that the different binning only slightly alters the best fit value of *N*.

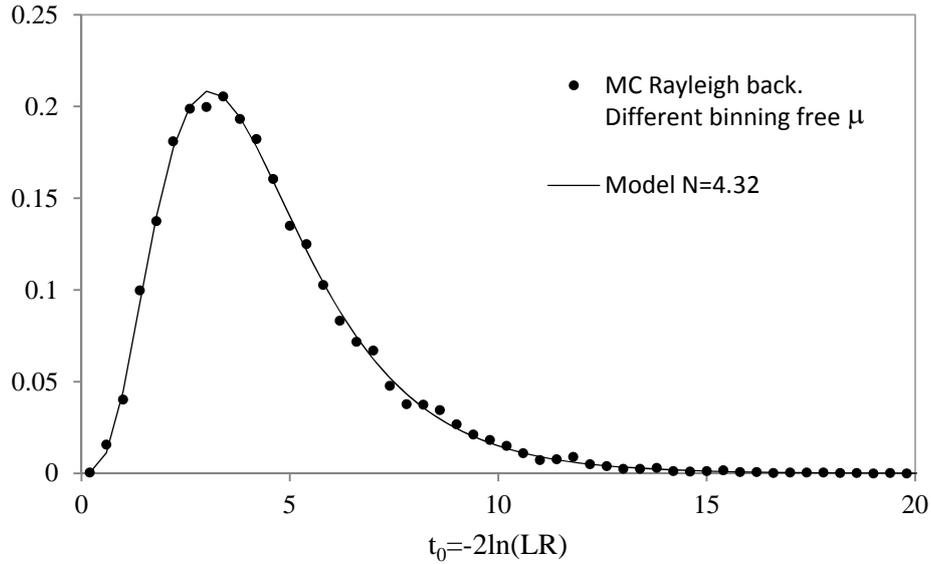

*Fig. 4 – Changing the binning does not substantially modify the outcomes of the MC test*

### 4.1.1 Other tests of the conjectured model

In order to bring additional evidence to the conjecture about the distribution of the $t_0$ statistics, some further specific tests have been performed.

#### 4.1.1.1 Distribution of $t_0$ in each "effective" search region

The first concerns the hypothesis that locally in each effective search region $t_0$ is $\chi^2_{s+1}$ distributed, where in the present example $s+1$ is equal to two degrees of freedom. In order to check this occurrence, one can conceive to plot the distribution of the maximum height in one of these effective region; in practice, upon each simulation cycle and the calculation of the corresponding $t_0$ values over the entire energy range of interest, the peak of $t_0$ occurring at the lowest energy is found (the lowest energy boundary is excluded for this test) and put in histogram. In this way it is as if the first effective "local" search region were tested to check the validity of the hypothesis of the local $\chi^2_2$ distribution.

The result of this evaluation is reported in Fig. 5. Apart from the first point, which can be regarded as an outlier, the rest of the plot in the figure confirms nicely the supposed local $\chi^2_2$ behavior of the test statistics $t_0$, bringing thus strong evidence in favor of the validity of the first "ingredient" of the model.

#### 4.1.1.2 Distribution at the edges or at any other fixed location in the allowed mass range

When the search of the signal is performed at a fixed location the asymptotic distribution predicted by the Wilks' theorem under the null hypothesis is expected to be valid, thus one should recover by MC a $\chi^2_1$ distribution. This is indeed the case in the example we are considering, as can be appreciated in the following Fig. 6, which displays the MC $t_0$ distribution for fixed mass (in this example just at the middle of the allowed range) and the corresponding $\chi_1^2(t_0)$ distribution: the agreement is surely remarkably good. By repeating the same calculation at the extremes of the permitted energy interval, it is found that at the upper edge the MC-model agreement is excellent as shown in Fig. 6, while at the lower edge is somehow less good. In conclusion, these outcomes confirm also the second "ingredient" of the conjecture, i.e. the $\chi_1^2$ distribution for fixed mass search, and hence in particular at the limits of the allowed energy span.

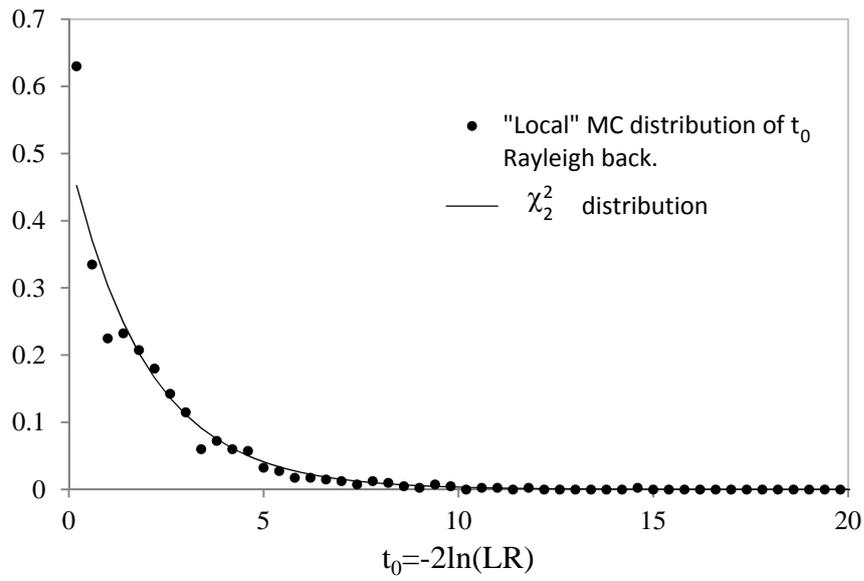

*Fig. 5 - Upon plotting the distribution of the lowest energy maximum of $t_0$, which amounts to test the behavior of the test statistics in the first of the "local" effective search regions, a good agreement with the hypothesized $\chi_2^2$ distribution is found.*

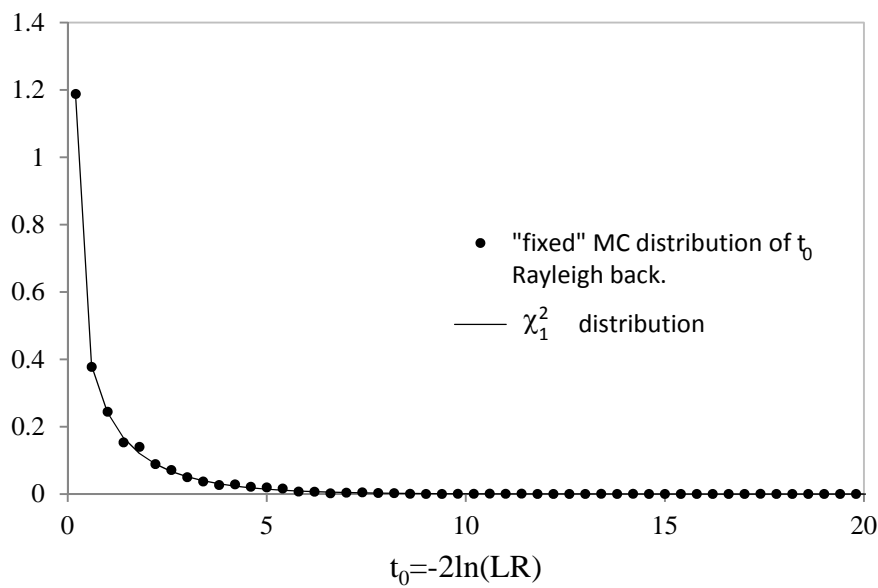

*Fig. 6 - When the discovery test statistics $t_0$ is evaluated for fixed mass the expected $\chi_1^2$ distribution is recovered, showing clearly the validity in this occurrence of the Wilks' theorem*

*4.1.1.3 Second ranked maximum*

One may wonder whether the conjecture that the maximum of the test statistics $t_0$ obeys the rules of order statistics is an occurrence limited to the highest peak only or is a manifestation of a more pervasive phenomenon affecting all the successive maxima of $t_0$. In the latter case one may expect that, for example, also the experimental MC distribution of the 2$^{nd}$ highest peak should be described by the prescriptions of order statistics.

In the specific situation of $N$ variables $\chi_2^2(t_0)$ distributed plus one variable $\chi_1^2(t_0)$ distributed the PDF of the height of the second highest peak can be written, following the methodology of order statistics, as

$$p(t_0) = \left( (N-1)\int_0^{t_0}\chi_1^2(h)dh \left(\int_0^{t_0}\chi_2^2(h)dh\right)^{N-2}\chi_2^2(t_0) + \chi_1^2(t_0)\left(\int_0^{t_0}\chi_2^2(h)dh\right)^{N-1}\right)N\int_{t_0}^{\infty}\chi_2^2(h)dh +$$

$$+ N\left(\int_0^{t_0}\chi_2^2(h)dh\right)^{N-1}\chi_2^2(t_0)\int_{t_0}^{\infty}\chi_1^2(h)dh \qquad (9)$$

The way in which the expression (9) is derived is based on considering one peak as the global highest and then writing the probability of the highest among the remaining peaks. More specifically, in the first term of the sum the factor on the right, e.g. $N\int_{t_0}^{\infty}\chi_2^2(h)dh$, corresponds to the probability that anyone of the $N\chi_2^2(t_0)$ variables is the highest peak, while the term on its left, within the brackets, expresses the probability of the highest among the residual $N-1$ $\chi_2^2(t_0)$ and the one $\chi_1^2(t_0)$ variables. Similarly, in the second term of the sum the factor on the left, e.g. $\int_{t_0}^{\infty}\chi_1^2(h)dh$, represents the probability that the highest variable is the one $\chi_1^2(t_0)$ distributed, while the three terms on its left express altogether the probability of the highest among the residual $N\chi_2^2(t_0)$ distributed variables.

To check whether this model is actually applicable, 10000 simulation cycles have been repeated and the height of the 2$^{nd}$ highest peak of the test statistics $t_0$ in each cycle has been recorded and put in a histogram, which is reported in Fig. 7.

Remarkably, the MC-model agreement is very good; it should be noted that it occurs for the value of $N = 5.16$ +/-0.044, thus different from the value 4.48 which ensured the MC-model agreement in the same case for the highest peak (see Fig. 1 and 2).

Therefore, this test indicates that the connection between the distribution of the maximum of the $t_0$ test statistics and the order statistics is likely a more general phenomenon, not restricted only to the highest maximum, though presumably not governed by a same $N$ parameter, as suggested by the present outcome.

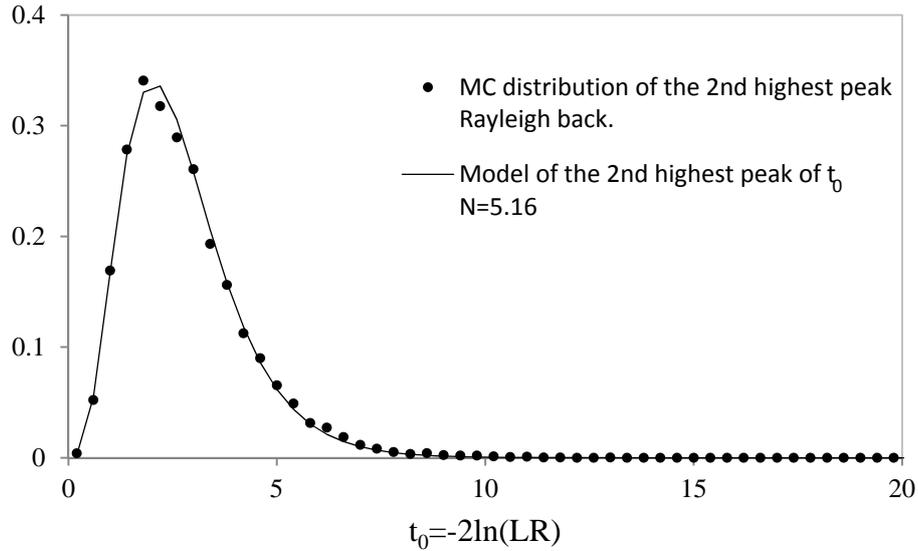

*Fig. 7 –The distribution of the second ranked peak of the test statistic $t_0$. It features a good match with the corresponding order statistics inspired model, see expression (9) in the text*

*4.2 Exponential background shape*

The $t_0$ distribution as stemming from the second toy model based on an exponential-like background is plotted in Fig. 8.

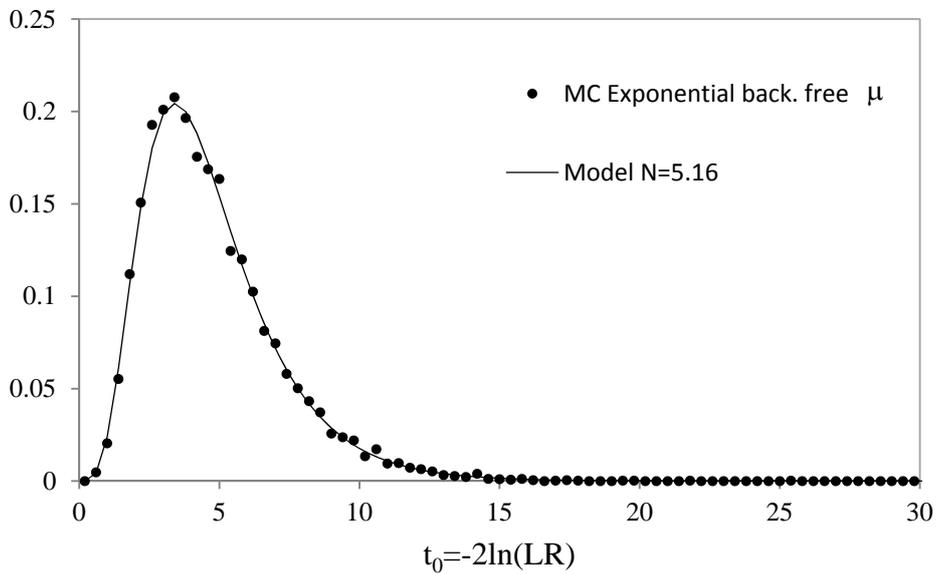

*Fig. 8 - Distribution of the $t_0$ test statistics for exponential background and with the signal strength parameter free to assume positive or negative values*

As in the previous Rayleigh example, the agreement with the conjectured model is extremely good. It has to be noted that in the MC generation procedure, as variant with respect to the Rayleigh based toy MC, the global number of background events has been generated as variable in each simulation cycles, while was kept fixed in the Rayleigh example. However, this procedural difference does not lead to change the outcomes of the test. *N* is fitted equal to 5.16+/- 0.055.

*4.2.1 $t_0$ distribution of the first "local" maximum and at fixed energy*

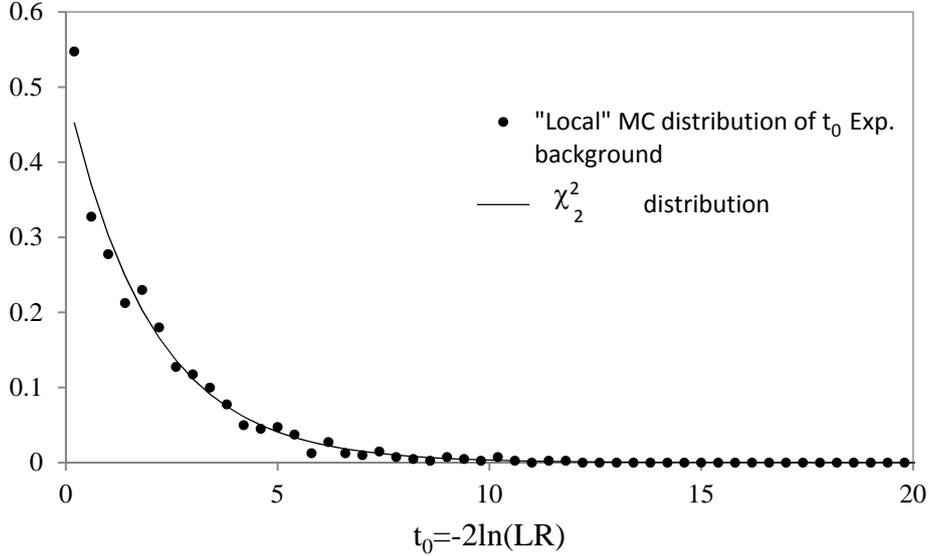

*Fig. 9 – Similarly to what found in Fig. 5, the distribution of the lowest energy $t_0$ maximum, corresponding to the behavior of the test statistics in the first of the "local" effective search regions, exhibits a good agreement with the expected $\chi^2_2$ distribution also when the background is modeled with an exponential function.*

Also in the present case of exponential background the result in Fig. 8 can be usefully supported by a simulation of the other two key aspects of the conjecture, e.g. the distribution of the local maximum of $t_0$ and the distribution obtained at fixed mass location.

The former MC distribution is reported in Fig. 9 together with the $\chi^2_2$ function: the agreement is definitively very good, as it was in the previous Rayleigh background example.

The latter MC distribution, which for the purpose of this example has been evaluated at the mass location 0.5, in the middle of the assumed mass range, is visible in Fig. 10, accompanied by the relevant model, i.e. the $\chi^2_1$ function.

The correspondence between the distribution stemmed from the MC and the theoretically expected model is excellent.

Therefore, it can be concluded that also when the background is exponentially distributed the agreement of the conjectured model, as well as of its basic ingredients, with the MC output is convincing. One may therefore speculate that the specific form assumed by the background does not matter to ensure the validity of the model (7), as long as the condition outlined in [6] to get the bound of the tail distribution is satisfied, i.e. that $t_0$ is $\chi^2_1$ distributed for fixed *E*.

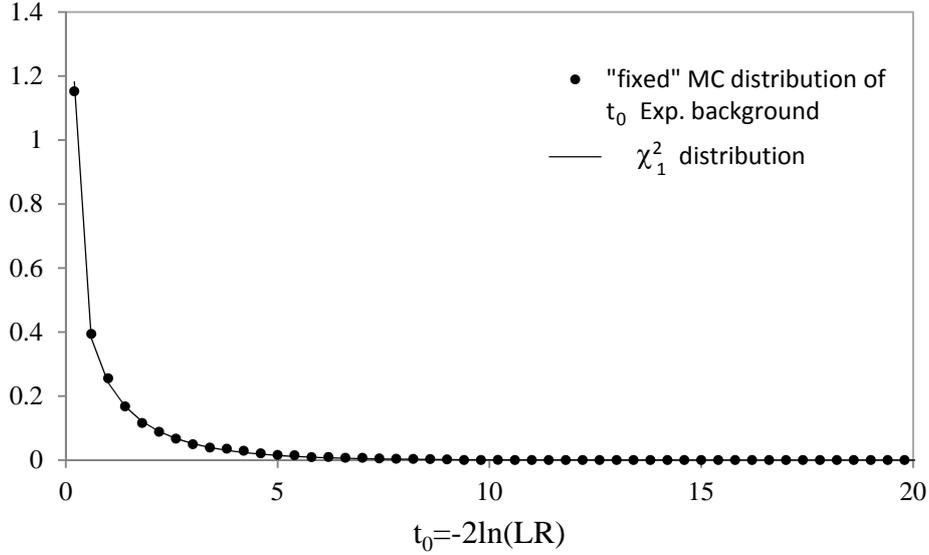

*Fig. 10 – The MC test at fixed mass for exponential background reproduces the same result of Fig. 6, i.e. a good agreement with the $\chi_1^2$ distribution, witnessing again that we are in the domain of applicability of the Wilks' theorem*

## 5. Numerical results of the toy models when considering the restriction of the signal strength parameter to be positive

The discussion at the end of the last paragraph implicitly assumes that the freedom allowed to the strength parameter to endow positive or negative values is pivotal in accomplishing the nice confirmation of the model got so far. It is in that occurrence, indeed, that the $\chi_1^2$ distribution for fixed *E* is recovered.

However, when looking for a bump on top of an invariant mass distribution is quite unnatural to admit for it a negative value. Rather, one would like to test the background-only hypothesis against the alternatives of positive-only signals. This task can be accomplished in two different ways, either simply restricting $\mu$ to be greater than 0 in the maximization of the denominator of the profile likelihood ratio, or by continuing to allow it to be positive or negative, but in the occurrence that a negative estimate $\hat{\mu}$ is obtained then $t_0$ is forced to 0.

The former prescription has clearly the drawback that the Wilks' conditions are violated also for the $\mu$ parameter, and thus we should not expect any more a $\chi_1^2$ distribution when $t_0$ is computed at fixed mass.

The latter instead, as shown in [14], ensures that the $t_0$ distribution for fixed mass will obey a so called ½ $\chi_1^2$, that is a distribution formed by a Dirac delta centered at 0, with area ½, and for t>0 by a usual $\chi_1^2$ function, but divided by 2.

The numerical results of both cases are examined in the following, to check how the findings of the previous paragraph get modified.

*5.1 Signal strength parameter restricted to be positive*

*5.1.1 Rayleigh background*

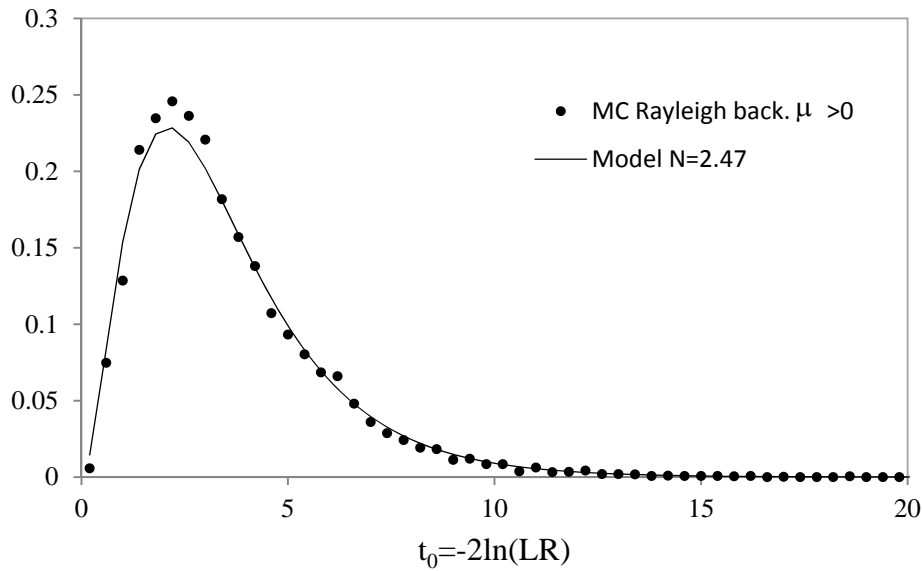

*Figure 11 – Distribution of the $t_0$ test statistics for Rayleigh background and with the signal strength parameter restricted to assume positive-only values.*

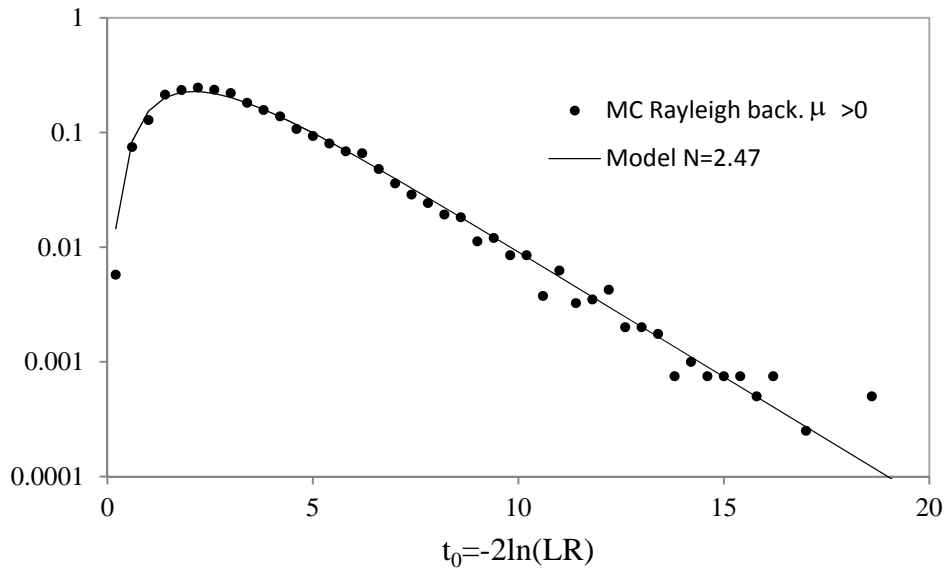

*Fig. 12 - The log plot of the same distribution of Fig. 11 demonstrates a reasonable match of the MC tail with the conjectured model, despite the non perfect overall model-MC correspondence*

Fig. 11 displays the simulation result for the Rayleigh type background, obtained while restricting the signal strength parameter to be positive. Two circumstances can be immediately noted, that the agreement with the model is only approximate, though still acceptable, the largest deviation occurring at the peak, and that the best fit parameter *N* has a value lower than in the of absence of restriction on the variability of $\mu$, e.g. 2.47 +/- 0.034 vs. the previous 4.32 +/- 0.05. In order to have some insight on the latter fact, if one plots in several simulation cycles the $t_0$ statistics as function of the energy, it results indeed that in average $t_0$ features less peaks in the present case than in the previous condition. Therefore, it is as if the number of fluctuating regions would decrease as a consequence of the reduced degree of variability of the strength parameter.

The log scale plot in Fig. 12 demonstrates a reasonable agreement at the level of the tail, despite the non perfect overall correspondence between the model and the MC data.

Also in this case $N$ can be estimated from the low statistic MC performed with only 100 simulations, obtaining as best estimate 2.86 +/- 0.34, thus again in reasonable agreement with the estimate from the 10000 simulation cycles.

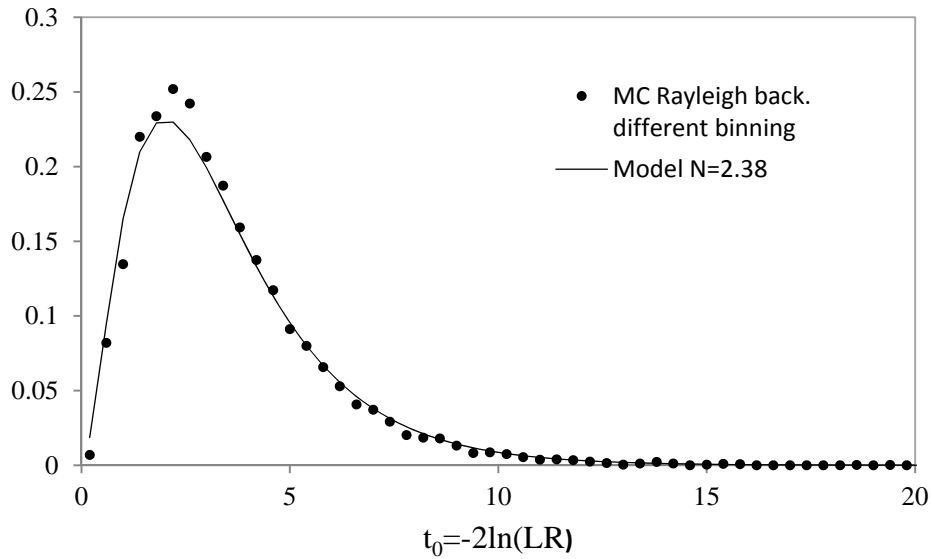

*Fig. 13 – As in Fig. 4, a modification of the binning does not substantially modify the outcomes of the MC test also when $\mu$ is restricted to be positive*

Fig. 13 reports the result of the same binning test performed in §4, increasing the bin size from 1 to 3. Very similar are also the conclusions, in the sense that the type of approximate comparison with the model remain unchanged, with a modest modification of $N$ that now results $N$= 2.38 +/- 0.035.

*5.1.1.1 – Evaluation of the building blocks of the conjecture when $\mu$ is constrained to be only positive*

We anticipated above that, since in this case $\mu$ is at the boundary of the allowed region, we do not expect the distribution of $t_0$ for fixed mass to follow the $\chi_1^2$ distribution. The MC result confirms this circumstance, as can be appreciated in Fig 14.

Similarly, while repeating in the present case the test to plot the distribution of the first "local" peak, the good agreement with the $\chi_2^2$ distribution reported in Fig. 5 and 9 is now lost.

This can be noted in Fig. 15, that shows a small, but significant, deviation from the $\chi_2^2$ function of the MC distribution of the first (first along the energy range, as usual) local maximum of $t_0$, with a sizable fraction of the MC points lying below the theoretical curve.

Since actually a $\chi_2^2$ distribution is simply an exponential function, i.e. specifically ½·exp(-½·x), we can easily play with it to get a better agreement with the MC distribution in Fig. 15, which can be obtained through a slight modification of the value of the normalization constant 2. Indeed, a best match is obtained by replacing 2 with 1.82, ignoring in any case the first point in the distribution, since it is really an outlier.

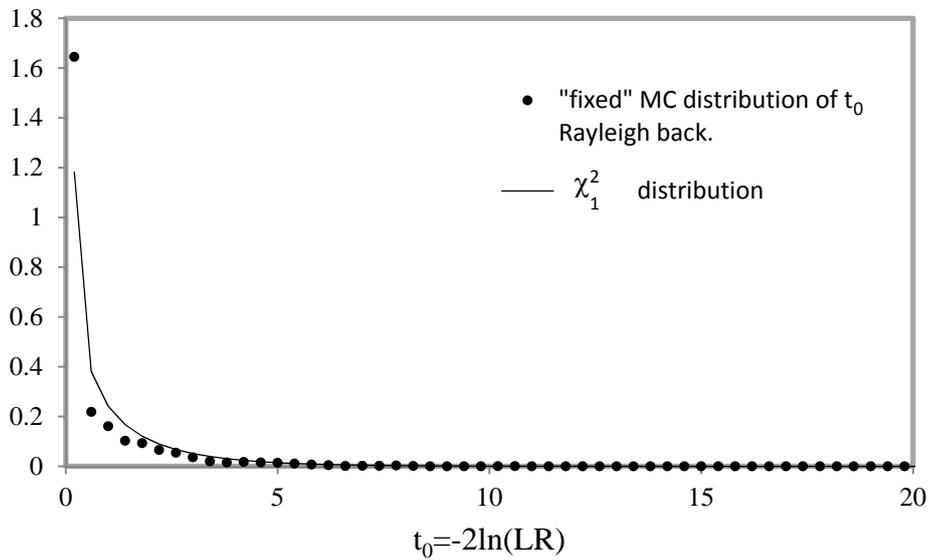

Fig. 14 - *When the discovery test statistics $t_0$ is MC evaluated for fixed mass the expected $\chi_1^2$ distribution is not recovered, thus confirming the violation of the conditions required for the validity of the Wilks' theorem*

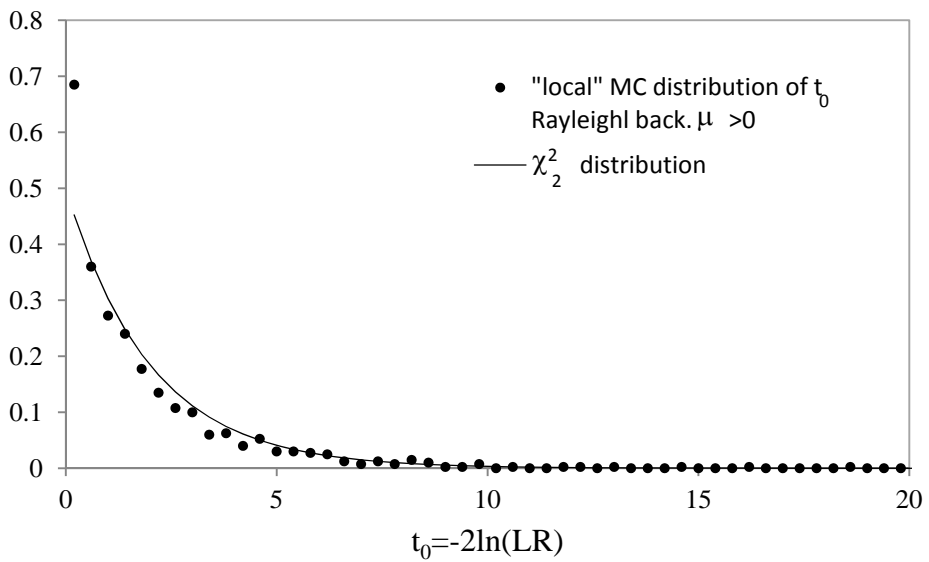

Fig. 15 - T*he distribution of the $t_0$ peak in the first of the "effective" local search regions exhibits a small but significant deviation from the hypothesized $\chi_2^2$ distribution, with several of the MC points below the theoretical curve*

The two results in Fig. 14 and 15 explain the not perfect agreement observed in Fig. 13 between the model and the MC of $t_0$ over the whole mass range, since both pre-requisites of a "local" $\chi_2^2$ distribution and of a fixed mass $\chi_1^2$ are not satisfied; furthermore, this observation suggests a way to alleviate the discrepancy, which consists in replacing in the model (7) the exact $\chi_2^2$ and $\chi_1^2$ distributions with some modified versions more closely adhering to the respective MC outputs.

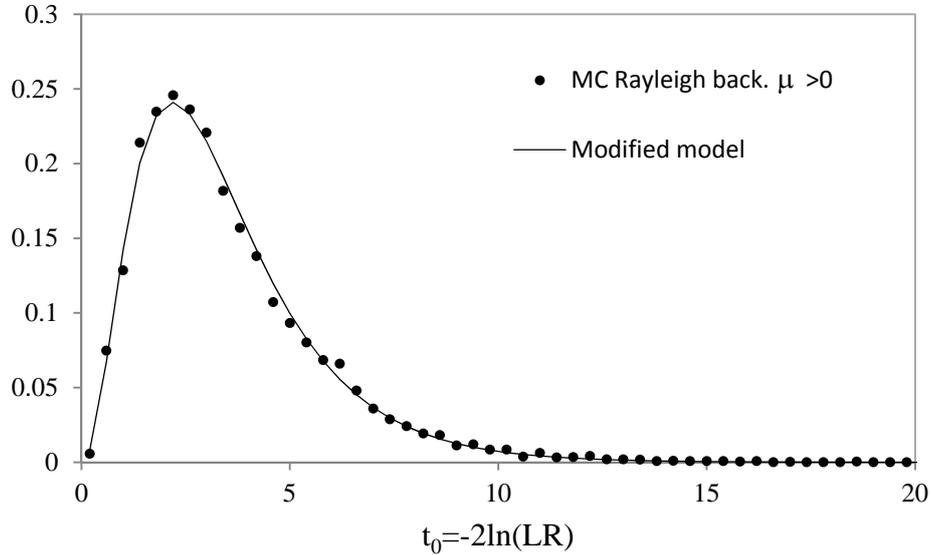

*Figure 16 – With respect to Fig. 13 a better MC-model agreement is obtained, in the case in which $\mu$ is restricted to be positive, by replacing in (7) the $\chi_2^2$ with a function more in agreement with the simulated local distribution*

Actually, an attempt has been tried consisting only in replacing the $\chi_2^2$ distribution with its modified version obtained replacing 2 with 1.82. The output of this test is reported in Fig. 16, showing the successful accomplishment of the goal to get a better agreement of the MC with the (modified) model. The fact that the modification of $\chi_2^2$ only is already enough to recover a good model-MC agreement can be explained by inspecting (7), and noting that the weight of $\chi_2^2$ is numerically overwhelming with respect to that of $\chi_1^2$ in determining its behavior. Also, the positive outcome of this attempt brings additional evidence of the validity of the conjecture, since it has been derived following its prescriptions, just modifying one ingredient, i.e. the ideal "local" $\chi_2^2$ distribution replaced with the actually MC observed "local" distribution.

*5.1.2 Exponential case*

The result of $t_0$ as stemming from the application of the maximization procedure in the exponential background toy model, by restricting $\mu$ to be positive, is reported in Fig. 17. Though not as excellent as in Fig. 8, the agreement between MC data and model, obtained for N=2.95 +/- 0.04, is remarkable, and definitely better than in the previous Rayleigh exemplification. Therefore, contrary to the paradigmatic case of § 4, it appears that when the condition $\mu$ >0 is imposed, which we know leads to violate the Wilks' conditions for fixed $E$, the degree of agreement with the model somehow depends upon the shape of the assumed background.

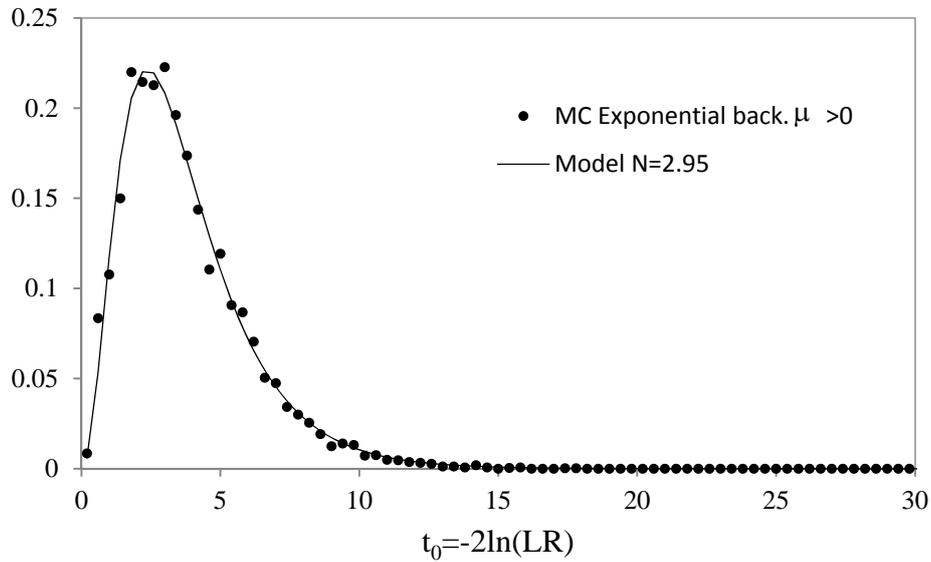

Fig. 17. - *Distribution of the $t_0$ test statistics for exponential background and with the signal strength parameter restricted to assume positive values*

*5.1.2.1 Evaluation of the building blocks of the conjecture when $\mu$ is constrained to be only positive*

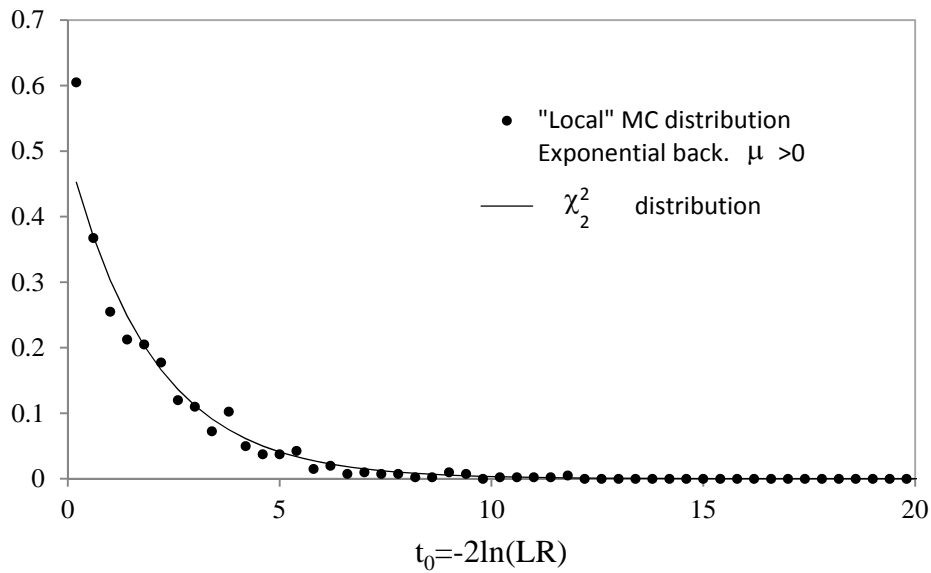

Fig. 18 – *Differently to what shown in Fig. 15, in the present case of exponential background the distribution of the lowest energy $t_0$ peak, corresponding to the behavior of the test statistics in the first of the "local" effective search regions, exhibits a reasonable agreement with the expected $\chi_2^2$ distribution*

The evaluation of the "building blocks" of the conjecture helps shed more light also on this specific example. The distribution of the first "local" peak, shown in Fig. 18, follows reasonably well $\chi_2^2$, not perfectly, but surely slightly better than in the previous Rayleigh example (see again Fig. 15 for comparison).

On the other hand, the $t_0$ distribution for fixed E as displayed in Fig. 19 disagrees with the $\chi_1^2$ function, of an amount similar to what also detected in the Rayleigh exemplification (Fig. 14), hence further reinforcing the indications of the previous test indicating that, numerically, the global agreement of $t_0$ with the model (7) does not depend much on it, but that, rather, is heavily dictated by the behavior of the distribution of the local peak.

*5.2 Forcing $t_0 =0$ if the signal strength parameter is found negative*

The alternative way to cope with the need to restrict our problem to a region physically admissible is to adopt the prescription of still allowing the strength parameter to assume positive or negative values throughout the maximization, but forcing $t_0$ to be equal to 0 if $\hat{\mu}$ is found negative. Interesting, the results do not change much with respect to the simple condition $\mu>0$, as shown in the next subparagraphs.

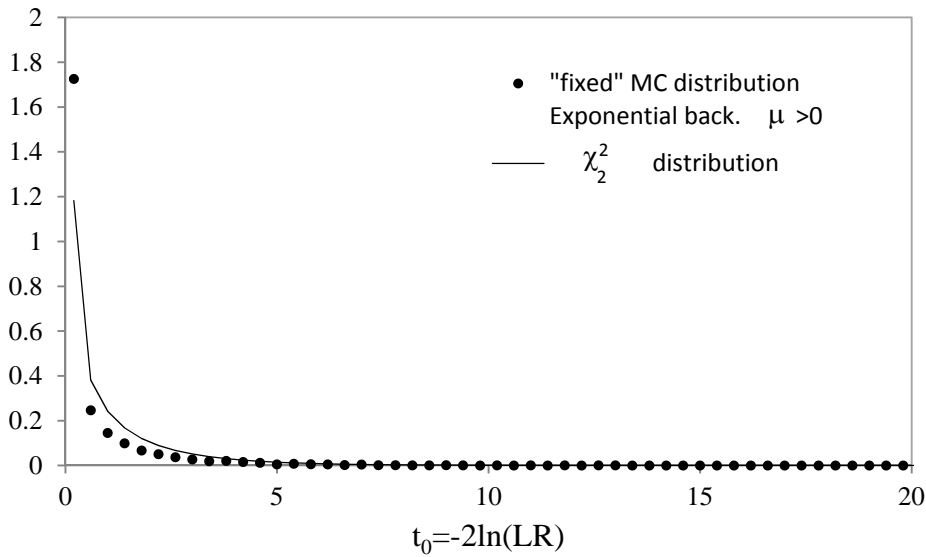

Fig. 19 – *As in the Rayleigh case, when the discovery test statistics $t_0$ is evaluated via MC at fixed mass the $\chi_1^2$ distribution is not recovered, thus confirming the violation of the conditions required for the validity of the Wilks' theorem caused by the restriction $\mu>0$*

*5.2.1 Rayleigh case*

By repeating the evaluation of the $t_0$ distribution in the Rayleigh case with the above prescription, the obtained MC output is shown in Fig. 20, displayed together with both the model and the same distribution found in the previous condition of allowing $\mu$ only positive, that is the distribution reported in Fig. 11. It clearly appears that the two prescriptions essentially originate the same result for $t_0$.

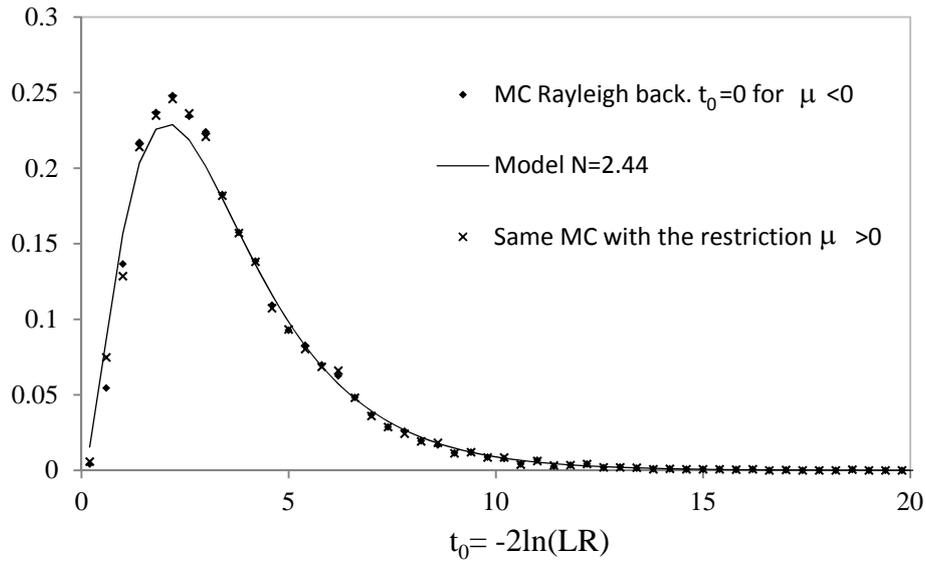

Fig. 20 - *Distributions of the $t_0$ test statistics for Rayleigh background, evaluated in both ways to cope with the positive restriction of the signal strength parameter. The two distributions are extremely similar.*

More insight comes from the usual simulation of the two key elements of the conjecture, the distributions at fixed mass and for the "local" search regions. The distribution at fixed mass is shown in Fig. 21, from which it can be inferred that, as expected, a good recovery of the $\chi_1^2$ distribution is obtained.

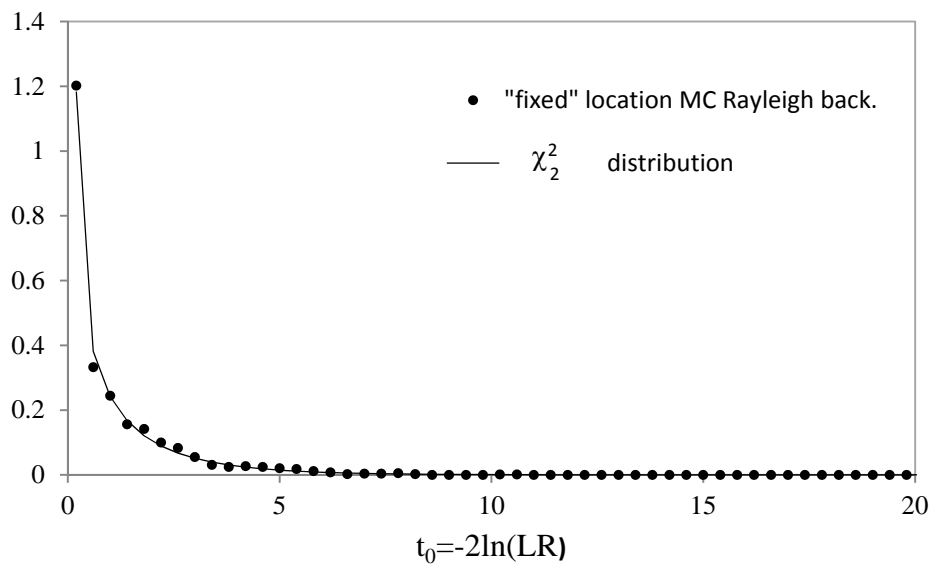

Fig. 21 – *The prescription to force $t_0=0$ when the signal strength parameter is fitted to a negative value recovers an ½ $\chi_1^2$ distribution. In the figure indeed half of the simulation results are plotted, those corresponding to positive outcomes of the signal strength parameter and they follow well the $\chi_1^2$ distribution. The other half outputs, not shown, would accumulate at zero, originating overall the ½ $\chi_1^2$ function*

It must be clarified that the MC distribution reported in Fig. 21 is built taking into account only half of the simulation results, since the other half would accumulate to 0, reproducing altogether the ½ $\chi_1^2$ distribution mentioned at the beginning of the paragraph.

However, when the distribution of the "local" peak is plotted ( as usual, the first peak encountered along the energy range is taken for this purpose) it comes out that it is essentially equal to the distribution reported in Fig. 15 for the case of $\mu$ restricted to be positive (and therefore due to this similarity is not reported here). This outcome explains why globally the $t_0$ distribution does not change toggling between the two prescriptions, as indicated by Fig. 20.

This result deserves some comments. One may have conceived that the recovery of some degree of regularity for the distribution at fixed mass through the ½ $\chi_1^2$ distribution could have been equivalent to satisfy the condition in [6] of $\chi_1^2$ distribution at fixed E, i.e. the condition that not only ensures the asymptotic behavior of the tail of $t_0$, but that the tests in §4 indicate as the pre-requisite for the full validity of the conjectured model (7). On the contrary, the MC result reported here shows that this is not the case, and in fact the global $t_0$ distribution does not change with respect to the previous simpler prescription in which one directly restrict $\mu$ to be greater than 0. In other words, the prescription used here which leads to obtain at a specified mass the ½ $\chi_1^2$ distribution is not equivalent to the condition of full validity of the Wilks' theorem when, allowing $\mu$ to be both positive or negative, the full $\chi_1^2$ distribution is recovered (from which the nice regular results of §4 stem).

*5.2.2 Exponential case*

The last considered case in these set of MC studies is the exponential background with the same prescription to force $t_0$=0 if $\mu$ is found negative. Actually, the achieved results are similar to those pertaining to the Rayleigh case.

First of all, the global $t_0$ distribution is practically indistinguishable from that in Fig. 17, corresponding to the $\mu$>0 restriction (and therefore not repeated here).

Furthermore, the $t_0$ local peak distribution is reported in Fig. 22, showing a reasonable agreement with the reference $\chi_2^2$ function.

Practically, the outcome in Fig. 22 is very similar to that in Fig 18, and reinforces the case that when the "local" peak distribution and the $\chi_2^2$ function do not deviate much each-other, then the agreement between the global $t_0$ distribution and the model is somehow ensured .

The last test concerns the evaluation performed at fixed mass, Fig. 23. By plotting the about half simulated cases in which $t_0$ was not forced to 0, since $\mu$ were found positive, we get the situation depicted in the figure, showing a good agreement with the $\chi_1^2$ function, similar to what illustrated in Fig. 21 for the Rayleigh case.

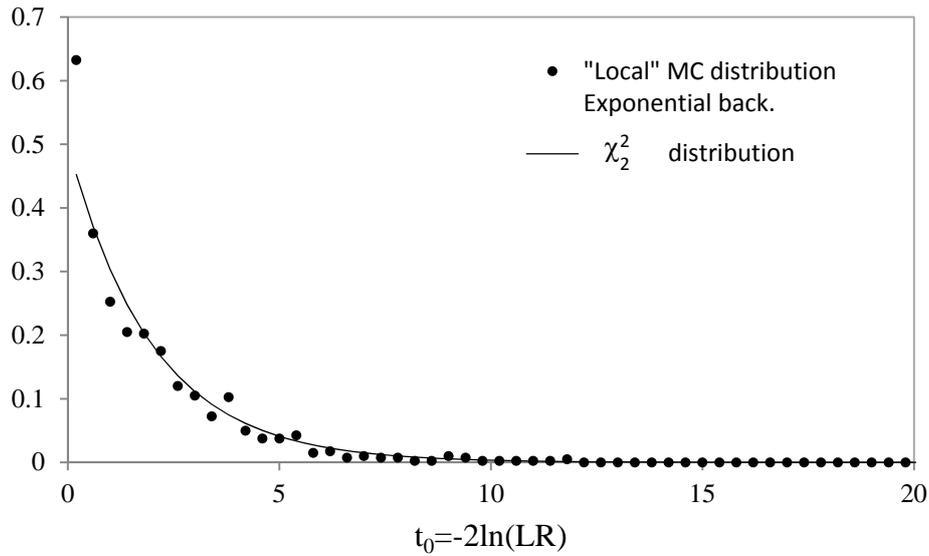

Fig. 22 – When the prescription $t_0=0$ is applied in the situation of exponential background, the distribution in the "local" effective search regions exhibits a reasonable agreement with the expected $\chi_2^2$ distribution. This output is extremely similar to that in Fig. 18, corresponding to the alternative prescription to force $\mu$ directly to be positive-only

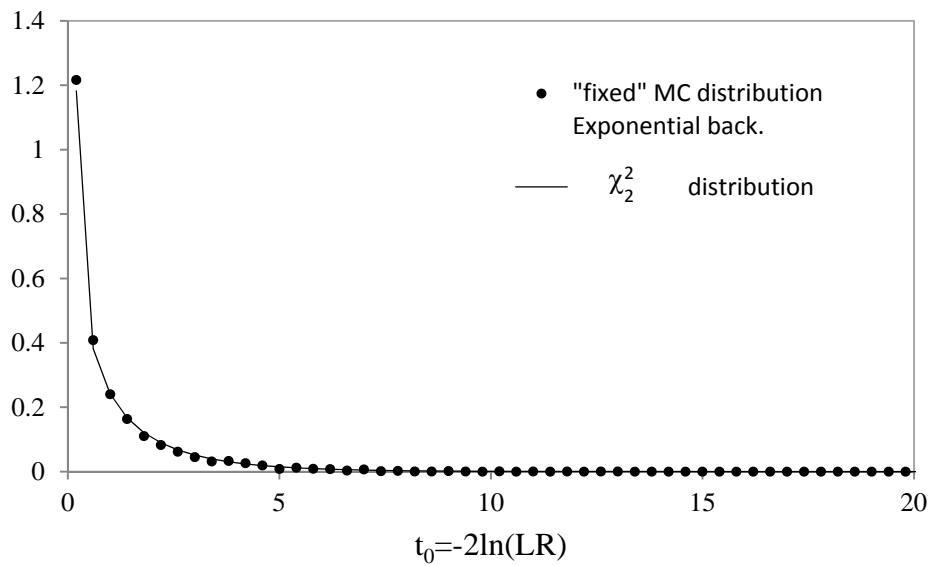

Fig. 23 - The prescription to force $t_0=0$ when the signal strength parameter is fitted to a negative value recovers an ½ $\chi_1^2$ distribution also when the background is of exponential type. As in Fig. 21, only half of the simulation results are considered in the plot, while the remaining half would accumulate at zero.

In conclusion, the outputs stemming from the exponential background case confirm that the two prescriptions to deal with the restriction of $\mu$ to positive values essentially have the same impact on the distribution of the discovery test statistics $t_0$.

# 6. Discussion

The set of tests performed above, in several different conditions, show that the conjectured functional form (7) for the distribution of the $t_0$ statistics is generally a good approximation of the true distribution; in particular, in all the examined variants it results that when the signal strength parameter $\mu$ is allowed to take on both positive or negative values throughout the maximization process of the denominator of the profile likelihood, then not only the MC distribution of $t_0$ over the considered energy range follows very precisely the model, but also the other two key ingredients, i.e. the distribution for fixed mass and the "local" distribution in each of the effective search regions, behave respectively as $\chi_2^2$ and $\chi_1^2$, as presumed in the conjecture itself.

That the distribution for fixed mass is $\chi_1^2$ distributed is actually expected from the Wilks' theorem, whose validity in this framework is ensured by the freedom allowed to $\mu$ to float between positive and negative values. Instead, the occurrence that the "local" distribution is actually $\chi_2^2$ is an important confirmation stemming from this MC study.

When there is no restriction on the variability of $\mu$, and therefore we are strictly in the condition of reference [6] that for fixed $E$ $t_0$ is $\chi_s^2$ distributed with $s$ degree of freedom (for simplicity $\chi_1^2$ in the concrete examples considered in this work) then the MC studies of § 4 strongly support the precise and accurate validity of the conjectured model; additional evidence in this sense is gained through the observation that the tail of the conjectured model of $t_0$ reproduces the asymptotic bound as derived in the same reference [6].

When, instead, $\mu$ is restricted to assume only physically possible positive values, either imposing directly this condition in the maximization associated with the profile likelihood ratio or through the indirect way of forcing $t_0=0$ if $\mu$ is found negative, then the correspondence between the model and the MC tests seems to be still reasonable, but not as precise as in the previous case, and somehow depending upon the shape of the background.

The key point to interpret this outcome is that when $\mu$ is not allowed to take on negative values the validity of the Wilks' theorem is no more ensured and thus the pre-requisite to have $t_0$ $\chi_1^2$ distributed at fixed $E$ is violated, a condition which the tests in § 4 showed to be of paramount relevance to extend the asymptotic bound of the tail derived in [6] to the conjecture about the entire shape of $t_0$.

The MC results lead to observe that in this situation the degree of correspondence to the model depends upon the actual shape of the assumed background, and that the "local" distribution of $t_0$ in the effective search regions is the key factor to guarantee, on a practical basis, the degree of accuracy with which the model reproduces the MC output. This is also proven by the fact that if in the model the $\chi_2^2$ function is replaced by a function more closely reproducing the actual MC distribution stemming from the "local" test, the global correspondence of the conjecture with the MC data is definitively improved.

These conclusions do not change substantially in the other prescription to deal with the negative value of $\mu$, despite the fact that in this way for fixed mass a certain form of regularity is recovered through the ½ $\chi_1^2$ function, but this appears not enough to recover the very regular results of § 4. In general, instead, the results associated with the previous prescription to strictly impose $\mu>0$ are reproduced.

**Acknowledgements**

The author would like to thank Eilam Gross and Ofer Vitells for illuminating discussions and the organizers of the Phystat 2011 workshop, whose stimulating atmosphere triggered this work.